\documentstyle[12pt]{article}
\begin{document}
\begin{titlepage}
\begin{flushright}   NSF-ITP-93-41; McGill/93-04; UMDGR-93-179\\
                     hep-th/9305016\\
\end{flushright}
\begin{center}
   \vskip 3em
  {\LARGE  Entropy of Lovelock Black Holes}    % title
  \vskip 1.5em
  {\large
    Ted Jacobson\footnote{jacobson@umdhep}\\[.5em]}
\em{ Institute for Theoretical Physics, University of California,
Santa Barbara, CA 93106\\
%and\\
Department of Physics, University of Maryland, College Park,
MD 20742-4111}\\[.7em]
 {\large
    Robert C. Myers\footnote{rcm@hep.physics.mcgill.ca}\\[.5em] }
\em{ Department of Physics, McGill University, Montreal, Quebec,
Canada H3A-2T8}
\end{center}
\vskip 1em
\begin{abstract}
A general formula for the entropy of
stationary black holes in Lovelock gravity
theories is obtained by integrating the first law of black hole
mechanics, which is derived by Hamiltonian methods.
The entropy is not simply one quarter
of the surface area of the horizon, but
also includes a sum of intrinsic curvature invariants integrated
over a
cross section of the horizon.
\end{abstract}
\end{titlepage}
\newpage                           %  Remove this line when making galleys

%=========================================================

The study of black hole thermodynamics is motivated
primarily by the hope of learning something about the nature of
quantum gravity.
Probing the limits of validity of the four laws of
black hole thermodynamics\cite{four} may provide one source of
insight in this context.
For instance, one would like to know
to what extent these laws are valid
when the back-reaction of quantum
field energy is taken into account.  Back-reaction
leads to the consideration of higher curvature interactions, which
arise from quantum field
renormalization\cite{field}.
This observation motivates
the investigation of black hole thermodynamics in
higher-curvature theories of gravity.  Another motivation
comes from string theory, where the low energy effective field
theory of gravity contains higher-curvature terms\cite{string}.

In this letter, we focus on a special class of higher-curvature
theories, called Lovelock gravity\cite{Lovelock}.
These theories are the most general
second-order gravity theories in higher dimensional spacetimes.
Higher dimensional spacetimes are of interest in several candidate
frameworks for unifying gravity with other interactions.
Moreover in higher dimensions with Lovelock theory, one can
explore the effect of higher curvature terms in black hole
thermodynamics without having to deal with complications that arise
in true higher derivative theories.
The Lagrangian density
for Lovelock gravity in a spacetime of dimension $D$
may be written
${\cal L}={\sum_{m=0}^{[D/2]}} c_m {\cal L}_m$ where
\begin{equation}
 {\cal L}_m(g)
=\frac{1}{2^m} \sqrt{-g}\;\;
\delta_{c_1d_1\cdots c_md_m}^{a_1b_1\cdots a_mb_m}\;\;
R_{a_1b_1}{}^{c_1d_1}\cdots R_{a_mb_m}{}^{c_md_m}
\label{lagrange}
\end{equation}
and ${\cal L}_0=\sqrt{-g}$\cite{footnote1}, and in the sum, $[D/2]$
indicates the integer part of $D/2$.
The $\delta$ symbol is a totally antisymmetric product of $2m$
Kronecker
deltas, normalized to take values $0$ and $\pm 1$.
Note ${\cal L}_1=\sqrt{-g}R$ by itself yields the
Einstein Lagrangian.  In general, ${\cal L}_m$ is the Euler class
for a $2m$-dimensional manifold.
Because of the antisymmetrization, no derivative appears at higher
than
second order in the equations of motion.
Static, spherically symmetric black hole solutions have
been found for Lovelock gravity\cite{hole,wilt}, and
given these solutions, one can examine to what extent the laws of
black hole thermodynamics apply\cite{jsz}.

Using the Euclidean approach\cite{Euclid},  it is clear
many of the essential features survive.
Given a specific stationary
solution, a Euclidean section
is obtained by analytic continuation.
Regularity of this section
requires identifying the Killing time
coordinate with a period $\beta=2\pi/\kappa$.
This section is regarded as a background in a periodic Euclidean path
integral, which is interpreted as giving a thermodynamic partition
function $Z$ with inverse temperature, $\beta$. In a semi-classical
approximation,
the Euclidean action $I(\beta,J,\cdots)$, regarded as a function of
$\beta$
and the extensive parameters (such as the angular momentum, $J$),
is identified with $-\ln Z$.
One defines the internal energy and entropy as
\begin{equation}
 U\equiv\frac{\partial I}{\partial\beta} \qquad\quad
S\equiv\beta\frac{\partial I}{\partial\beta}-I        \label{c}
\end{equation}
which then automatically satisfy the first law
\[
T\delta S=\delta U -
\left(T\frac{\partial I}{\partial J}\right) \delta J \;\; .
\]
This result reproduces the first law of black hole mechanics
if $U=M$, the black hole mass, and
$(T\,\partial I/\partial J)=\Omega$, the angular velocity
of the horizon. The latter relations appear to hold in
the present context ({\it i.e.,}\
higher curvatures and/or higher dimensions),
and it may be possible to prove them
by extending the methods of Ref.\cite{BrownYork}.
One might expect that $S$ coincides with one quarter the surface area
of the horizon\cite{footnote2} as in Einstein gravity,
but this identity fails in Lovelock
gravity\cite{wilt,jsz}, and other higher curvature
theories\cite{others}.

As described above, the Euclidean approach is applied to investigate
known stationary solutions.
While this yields $S$ as a function of the solution's parameters,
it provides no guidance as to any {\it geometric} significance of
the black hole entropy. It also fails to provide a definition for
$S$ that generalizes to an arbitrary, time-dependent horizon, without
which it is difficult to see how to even address the question of whether
a classical second law is obeyed in ``nonequilibrium" processes.
However, it is also possible to work with the Euclidean approach without
specific solutions in mind\cite{WhitYork,BrownYork}.
(Indeed, this has recently been done for static metrics in
Lovelock gravity with only $c_1$ and $c_2$ non-vanishing\cite{jsbw}.)
When applied in this more general fashion, it should be possible to avoid
the abovementioned drawbacks of the Euclidean approach.

Our alternate approach
is to find the {\it variation} $\delta S$ of the entropy
by deriving a general form of the first law of black hole mechanics
following the Hamiltonian
method introduced by Sudarsky and Wald\cite{SuWald}.
A priori, there is no guarantee that the resulting $\delta S$ is
the variation of a functional of the intrinsic geometry of the
horizon,
however, one finds that
it is for Lovelock gravity.
In fact, the entropy is given by an integral over a cross
section of the horizon, of an integrand which is similar
in form to the Lagrangian itself.

The method of Ref.\cite{SuWald} applies to {\it Killing horizons}.
A Killing horizon is a null hypersurface whose null generators are
tangent to a Killing field.
In {four}-dimensional Einstein gravity, Hawking proved
that a stationary black hole event horizon must be a Killing
horizon\cite{HE}.
This proof can not obviously be generalized to Lovelock gravity,
but the result is certainly true for all the known solutions
(which, however, are all static).

If $\chi^a$ is the Killing field which is null on the horizon,
the surface gravity $\kappa$ can be defined by
$\nabla_a(\chi^b\chi_b)=-2\kappa\chi_a$,  or equivalently,
$\chi^b\nabla_b\chi^a=\kappa\chi^a$.
If one assumes (as we will) that the null generators of the horizon
can be extended to the past as complete geodesics, three important
properties follow:
(i)~$\kappa$ is {\it constant} over the entire horizon \cite{RaczWald}.
(ii)~There exists a $(D$--2)-dimensional spacelike cross-section~$B$
of the horizon on which $\chi^a$ vanishes \cite{RaczWald}.
$B$~is called the {\it bifurcation surface}.
(iii)~The extrinsic curvature of $B$ vanishes \cite{CWald}.
A further property of Killing horizons that is important for our
purposes is that any two spacelike slices of the horizon
are isometric.

Let $\xi^a$ be the Killing field that is asymptotically a
pure time translation. Then by suitably rescaling $\chi^a$,
one is able to
reduce the Killing field $\chi^a-\xi^a$ to a rotation.
In higher dimensions,
there are $[(D-1)/2]$ commuting
Killing fields $\phi_{{\scriptscriptstyle(\alpha)}}^a$
generating
rotations in totally orthogonal planes\cite{merry}, and
so one has
\[
\chi^a=\xi^a+\Omega^{{\scriptscriptstyle(\alpha)}}
\phi_{{\scriptscriptstyle(\alpha)}}^a \;.
\]

Next, we briefly present the Hamiltonian description of
Lovelock gravity\cite{TZan}, which will be necessary in the following
derivation of the first law.  As usual after splitting space and time,
the dynamical variables
on a spatial surface $\Sigma$ are the spatial metric $h_{ab}$, and
its conjugate momentum $\pi^{ab}$\cite{footnote1}.
In the present case, $\pi^{ab}$ is a
complicated function of both the extrinsic and intrinsic curvatures
of $\Sigma$\cite{TZan}.
The normal and spatial components of the
time flow vector field $t^a$, the lapse $N$ and shift $N^a$,
are arbitrary Lagrange multipliers
in the Hamiltonian.
For asymptotically flat space,
the Hamiltonian has two parts $H=H_V+H_S$.
The volume term $H_V$ is a combination of constraints
${\cal H}_{\bot}$ and ${\cal H}_a$,
\[
H_V=\int_{\Sigma}d^{D-1}\!x\, (N{\cal H}_{\bot}+N^a{\cal H}_a)\;\;,
\]
and hence vanishes when evaluated for solutions of the field equations.
${\cal H}_a$ is the generator of spatial diffeomorphisms in $\Sigma$,
and so,
as in Einstein gravity,
${\cal H}_a=-2 D_b (\pi_a {}^b)$,
where $D_a$ is the covariant derivative compatible with
$h_{ab}$. ${\cal H}_{\bot}$ generates normal deformations of
$\Sigma$, and one finds
${\cal H}_{\bot}=\Sigma c_m {\cal H}_{\bot}^{(m)}$
with
\[
 {\cal H}_{\bot}^{(m)}= -\frac{\sqrt{h}}{2^m}\;
\bar{\delta}_{c_1d_1\cdots c_md_m}^{a_1b_1\cdots a_mb_m}\;\;
R_{a_1b_1}{}^{c_1d_1}\cdots R_{a_mb_m}{}^{c_md_m}
\]
where $R_{ab}{}^{cd}$ is the full $D$-dimensional curvature tensor.
Since $R_{ab}{}^{cd}$ is projected into $\Sigma$,
it can be replaced by
$\bar{R}_{ab}{}^{cd}+2K_{[a}{}^c K_{b]}{}^d$.
Here $\bar{R}_{ab}{}^{cd}$ is the curvature of $h_{ab}$, and
$K_{ab}$ is the extrinsic curvature of
$\Sigma$,
which is regarded
as a function of $h_{ab}$ and $\pi^{ab}$\cite{footnote3}.
The $\bar{\delta}$ symbol is the antisymmetric product of
projected Kronecker deltas,
$ \bar{\delta}_c^a=
\delta_c^a+n^a n_c$ where
$n^a$ is the unit normal to $\Sigma$.

The Hamiltonian also has an asymptotic boundary term, just as
in Einstein gravity\cite{RT}, $H_S=\oint_{\infty}
d^{D-2}\!x\,(NS_{\bot}+N^aS_a).$
$H_S$ is added to cancel surface terms
which arise by integrating by parts in $H_V$ when generating
the Hamiltonian
flow equations.  The precise value of $H_S$ depends on
the choice of $t^a=Nn^a+N^a.$  If asymptotically
$t^a\to\xi^a$, the time translation Killing field, then
$H_S=M$, the total mass of the solution.  If
$t^a\to\phi_{\scriptscriptstyle (\alpha)}^a$, a rotational Killing field,
then $H_S=-J_{\scriptscriptstyle (\alpha)}$, the associated angular momentum.

The derivation of the first law
comes from a judicious evaluation of the Hamiltonian\cite{SuWald}.
Let $h_{ab}$ and $\pi^{ab}$ describe a stationary black
hole solution,
and choose the ``time"
flow field to be the Killing field that is null on the horizon,
$t^a=\chi^a=\xi^a+
\Omega^{{\scriptscriptstyle(\alpha)}}
\phi_{{\scriptscriptstyle(\alpha)}}^a$.
Evaluate $H$ on a space-like slice $\Sigma$
that
extends from asymptotic infinity
to the horizon, intersecting the horizon at the bifurcation
surface, $B$.  With these choices one has
$H=M-\Omega^{\scriptscriptstyle (\alpha)}
J_{\scriptscriptstyle (\alpha)}.$
Let $(\delta h_{ab},\delta\pi^{ab})$ be a perturbation of the initial
solution to {\it any} nearby solution---not necessarily stationary.
With $t^a$ and $\Sigma$ fixed, one has
\[
\delta H= \delta M-\Omega^{(\alpha)} \delta J_{(\alpha)}
\]
On the other hand, one has Hamilton's equations
\[
\delta H=\int_{\Sigma} d^{D-1}\!x\left(
{\cal L}_{\chi}h_{ab}\,\delta\pi^{ab}
-{\cal L}_{\chi}\pi^{ab}\,\delta h_{ab}
\right)+\delta H_B
\]
The volume term vanishes because the flow is along a Killing field,
${\cal L}_{\chi}h_{ab}=0={\cal L}_{\chi}\pi^{ab}$.
Integration by parts is needed to produce
these volume terms. The surface terms at asymptotic
infinity are cancelled by the variation $\delta H_S$,
but one is left with surface terms, denoted $\delta H_B$, at the
inner boundary of $\Sigma$.

Since $\delta H_B$ is evaluated on the bifurcation surface $B$
where $t^a=\chi^a$ vanishes, any nonvanishing terms involve
derivatives of $t^a$.  Such terms only arise from the metric
variation of the curvatures in $N{\cal H}_{\bot}$.
Two integrations by parts arise from
$\delta \bar{R}_{ab}{}^{cd}=-2D_{[a}D^{[c}\delta h_{b]}{}^{d]}
+\bar{R}_{ab}{}^{e[c}\delta h_e{}^{d]}$,
where $\delta h_a{}^d=h^{db}\delta h_{ab}$.
The first
integration yields
new volume integrals involving $D_aN$. In these integrals,
the second integration by parts produces a boundary term at
$B$ involving $D_aN$.
The contribution from ${\cal H}_{\bot}^{(m)}$ is
\begin{equation}
\delta H_B^{(m)}=\frac{mc_m}{2^{m-1}}\oint_B
d^{D-2}\!x\,\left(\sqrt{\tilde{h}}\;v^{c_1}
D_{a_1}\!N\, \bar{\delta}_{c_1d_1c_2\cdots d_m}^{a_1b_1a_2\cdots b_m}
\delta h_{b_1}{}^{d_1}R_{a_2b_2}{}^{c_2d_2}\cdots
R_{a_mb_m}{}^{c_md_m}\right)
\label{hum}
\end{equation}
where $v^c$ is the unit normal
to the bifurcation surface $B$, pointing into $\Sigma$,
and $\tilde{h}$ is the determinant of $\tilde{h}_{ab}$,
the induced metric on $B$.
Now it is not hard to show that,
at $B$,
$D_aN=\kappa v_a$, where $\kappa$
is the surface gravity.
Since
$v_{a_1}v^{c_1} \bar{\delta}_{c_1d_1\cdots d_m}^{a_1b_1\cdots b_m}
= \tilde{\delta}_{d_1\cdots d_m}^{b_1\cdots b_m}$,
the antisymmetrized product of Kronecker deltas
$\tilde{\delta}_c^a=\delta_c^a+n^a n_c-v^av_c$
projected into $B$,
all remaining tensors are fully projected into $B$.
The projected $\delta h_b{}^d$ yields
$\delta\tilde{h}_b{}^d\equiv\tilde{h}^{de}\delta\tilde{h}_{be}$.
Since the extrinsic curvature of the bifurcation surface $B$
vanishes, projecting $R_{ab}{}^{cd}$  yields the intrinsic
curvature $\tilde{R}_{ab}{}^{cd}$ of $\tilde{h}_{ab}$.

Thus far variations of the $K_{ab}$ terms
in $R_{ab}{}^{cd}$
were neglected. Regarded
as a function of $\pi^{ab}$ and $h_{ab}$, $K_{ab}$ involves
$\bar{R}_{ab}{}^{cd}$ and so could produce extra boundary
contributions.
These extra terms always vanish
however,
since they contain at least
one other factor of $K_{ab}$ which vanishes when projected into $B$.
Since $\kappa$ is a constant on $B$, Eq.~(\ref{hum}) thus becomes
\[
\delta H_B^{(m)} =\frac{mc_m\kappa}{2^{m-1}}\;
\oint_B
d^{D-2}\!x\left(\sqrt{\tilde{h}}
\tilde{\delta}_{d_1\cdots c_md_m}^{b_1\cdots a_mb_m}\;\delta
\tilde{h}_{b_1}{}^{d_1}\cdots \tilde{R}_{a_mb_m}{}^{c_md_m}\right)\ .
\]
Extracting the temperature, $\kappa/2\pi$,
the contribution $\delta S^{(m)}\equiv(2\pi/\kappa)\delta H_B^{(m)}$
to the variation in the entropy thus depends
only on the intrinsic geometry of $B$.
The next step is to recognize that $\delta S^{(m)}$
is in fact the variation of an intrinsic geometric quantity
on $B$, $S^{(m)}=4\pi mc_m
\oint_B d^{D-2}\!x\,{\cal L}_{m-1}(\tilde{h})$
with ${\cal L}_{m}$ as defined in Eq.(\ref{lagrange}).

Collecting all of the contributions to the entropy, our final
result is the first law of black hole mechanics in Lovelock gravity,
$\delta M-\Omega^{\scriptscriptstyle (\alpha)}
\delta J_{\scriptscriptstyle (\alpha)}=
(\kappa/2\pi)
\delta S$ where
\begin{equation}
S=\sum_{m=1}^{[D/2]} 4\pi m\, c_m \oint d^{D-2}\!x \; {\cal L}_{m-1}
(\tilde{h})                                    \label{refinir}
\end{equation}
Our derivation of the first law relied heavily
on the fact that the variation of the entropy
is evaluated on the bifurcation surface $B$.
Nevertheless, the integrated expression (\ref{refinir})
for the entropy of a stationary black hole
can be evaluated on {\it any} spacelike slice of the
Killing horizon, since all such slices are isometric.
Note that with $c_1=1/16\pi G$,
as appropriate for the
Einstein action, the first term in the sum is simply $A/(4G)$, where
$A$ is the surface area of the horizon.
It is remarkable that the entropy (\ref{refinir}) is identical
in form to the original action, evaluated in a (Riemannian) space
of dimension $D-2$, with $c_{m-1}$ replaced by $4\pi mc_m$

For even dimensions,
we have included a non-trivial
integration constant in Eq.~(\ref{refinir}).  This constant is the
contribution of ${\cal L}_{D/2-1}$, which yields the Euler
constant of the cross
section of the horizon.  In four dimensions, this constant is fixed for
all stationary black holes since the horizon must have the topology of
$S^2$$\times$$R.$\cite{HE}
In $D$$>$4, the horizon topology is not unique,
so this topological entropy may have more significance.
With this choice for integration constant
one easily confirms that for the known
black hole solutions\cite{hole}, Eq.~(\ref{refinir}) produces the same
result as Eq.~(\ref{c}) in the Euclidean approach\cite{jsz}
(provided, of course, that $c_{D/2}$ is identical for both approaches).

The present derivation is easily extended to include Maxwell or Yang-Mills
fields\cite{SuWald}. In the case of charged black holes, the first law
becomes $\delta M-\Phi\delta Q-\Omega^{\scriptscriptstyle (\alpha)} \delta
J_{\scriptscriptstyle (\alpha)}
=(\kappa/2\pi) \delta S,$ where $\Phi$ is the potential difference
between the horizon and infinity, and $S$ is still given by
Eq.~(\ref{refinir}). Again, Eq.'s (\ref{refinir}) and (\ref{c}) yield
the identical results when applied to the known charged black hole
solutions\cite{wilt}.

Our results extend the
framework of black hole thermodynamics in a natural
way to Lovelock gravity. The zeroth law
({\it i.e.,} the constancy of $\kappa$) holds modulo the
assumption that the horizon generators are geodesically complete
(or, if the black hole forms from collapse, that the horizon
generators of the stationary solution which it approaches
have this property).
The first law defines an entropy that is localized in the
intrinsic geometry of the horizon and is valid for all
stationary black holes, at least provided the event horizon is a
Killing horizon.  For example, Eq.~(\ref{refinir}) applies for
rotating
black hole solutions, which are as yet unknown.
If all the coefficients $c_m$ are of the same order in units of some
common length scale, then the area term
($m=1$) dominates the entropy
for black holes that are much larger than that scale. The
entropy of small black holes can be negative, but for fixed
couplings $c_m$
it is bounded below
for the known solutions\cite{jsz}.
Such negative values are thermodynamically benign,
since only changes of the entropy occur in the first law.

The second law for quasi-stationary
processes follows from the
first law, provided the quantity
$\delta M-\Omega^{\scriptscriptstyle (\alpha)}
\delta J_{\scriptscriptstyle (\alpha)}$
is positive. This is the case for fluxes of
positive energy matter\cite{Carter}.
To control the sign for fluxes including gravitational energy
would require a positive energy theorem for Lovelock gravity,
which has not been established.\cite{footnote4}
(Negative energy states would probably
make the theory unstable however, in which case
a failure of the second law would be of less significance.)
Alternatively, the second law in Einstein gravity can be derived from
Hawking's area theorem\cite{HE} which shows that,
even in non-quasi-stationary processes,
the area (and therefore the entropy) will never decrease provided
the Ricci tensor (or, using Einstein's equation, the matter energy
tensor)
obeys the null energy condition.
In Lovelock gravity the expression (\ref{refinir}) for the entropy is
at least meaningful on a slice of a time-dependent horizon, so one
might hope to prove a non-equilibrium second law by
similarly following the
evolution of the horizon. This remains an interesting problem
for future work.
In this connection curiously, note that with $c_{D/2}>0$ the topological
term in Eq.~(\ref{refinir}) could lead to violations of the second law
when two black holes coalesce, even in four dimensions.

When quantum fields are included, negative energy can be transferred
to the black hole, as in Hawking evaporation\cite{HawkRad} or
``mining"\cite{UnWald} processes. The interesting question is then
whether the {\it generalized} second law (GSL)
($\delta(S_{BH}+S_{outside})\ge 0$)
holds.
Validity of the {\it classical} second law for $S_{BH}$ alone
would seem to be a prerequisite for validity of the GSL.
Provided Lovelock gravity turns out to satisfy the classical second
law, it seems to be
on the same footing with Einstein gravity as far as the
GSL is concerned.
For instance,
Zurek's argument\cite{Zurek} that entropy increases
during Hawking evaporation uses only the first law, and not the form
of the black hole entropy, so it will carry over to Lovelock gravity.
More general arguments\cite{gsl},
to the extent that they are valid,
also seem to carry over.
However, even if valid, these arguments
apply only to quasi-stationary processes, so the validity of
the GSL remains an important open problem.

Finally, Lovelock gravity is only a very special case of possible
generally covariant gravity theories. The fact that the entropy
is not one quarter the surface area, is already known in many other
examples of theories with higher curvature interactions\cite{others}.
If one derives a first law via the method of
Sudarsky and Wald as we have done here, the
variation of the entropy will
again
be given as an integral over the bifurcation surface of the
horizon.
However, this expression will not in general depend only on the
intrinsic geometry of the horizon, and it is not even clear that it
will be the variation of some quantity defined locally at the
horizon (although, from the Euclidean approach, Ref.~\cite{jsbw}
argues that the entropy is {\it always}
a quantity defined locally at the horizon).
In fact, the particular properties of Lovelock
gravity played a crucial role in our calculations establishing the
local and intrinsic nature of the entropy.  Whichever way it turns out
in general higher curvature theories, one of the authors of this
paper will enjoy a free dinner at a restaurant of his choice.

\vskip 3em
We would like to acknowledge useful discussions with D.~Brill,
D.~Brown, P.~Chrusciel, S.~Deser, G.~Horowitz, G.~Kang,
I.~Moss, J.Z.~Simon, A.~Strominger, R.~Wald, and J.~York. R.C.M.~thanks
the Institute for Theoretical Physics at UCSB for their hospitality
during this work. R.C.M.~was supported by NSERC of Canada, and Fonds
FCAR du
Qu\'{e}bec.  T.J.~was supported by NSF grant PHY91-12240.
Research at the ITP, UCSB was supported by NSF Grant PHY89-04035.

%==================================================================


\begin{thebibliography}{10}
\bibitem%[footnote1]
 {footnote1}We set $c_0$, the cosmological
constant, to zero, and consider only asymptotically flat
spaces in the following.  The spacetime signature is
taken as ($-$ $+$ $\cdots$ $+$). Throughout the paper
lower case latin letters are employed for all tensor
indices, including those on ``spatial" tensors. That is,
spatial tensors are just regarded as spacetime tensors which
vanish when any index is contracted with the normal(s) to the
spatial submanifold.
\bibitem%[footnote2]
 {footnote2}In $D$ dimensions, a cross section of the
horizon is a $(D$--2)-dimensional space, and so the surface area refers
to the volume of such a slice.
\bibitem
 {footnote3}In fact, one cannot in general solve uniquely for
$K_{ab}$ in terms of $\pi^{ab}$ and $h_{ab}$\cite{TZan},
but this ambiguity does not enter into our analysis.
\end{thebibliography}
\end{document}